\documentclass[prl,superscriptaddress,preprint,amsmath,amssymb]{revtex4}
\usepackage{graphicx}
\usepackage{indentfirst}
\usepackage{psfrag}
\usepackage{epsfig}
\usepackage{amsmath}
\usepackage{amssymb}
\usepackage{bm}

\begin{document}


\title{Optical forces from an evanescent wave on a  magnetodielectric small particle}


\author{M. Nieto-Vesperinas, $^{1,*}$ and J. J. Saenz, $^{2,* *}$}

\address{$^1$Instituto de Ciencia de Materiales de Madrid, Consejo Superior
de Investigaciones Cientificas, Campus de Cantoblanco, Madrid 28049,
Spain.
\\
$^2$Departamento de F\'{\i}sica de la Materia Condensada,
Universidad Aut\'{o}noma de Madrid, 28049 Madrid, Spain, and \\
Donostia International Physics Center (DIPC), Paseo Manuel
Lardizabal 4, 20018 Donostia-San Sebastian, Spain.\\
$^*$mnieto@icmm.csic.es \, \, $^{* *}$juanjo.saenz@uam.es}

\begin{abstract}We report the first study on the optical
force exerted by an evanescent wave on a small sphere with both
electric and magnetic response to the incident field, immersed in an
arbitrary non-dissipative medium. New expressions and effects from
their gradient, radiation pressure, and curl components are obtained
due to the particle induced electric and magnetic dipoles, as well
as to their mutual interaction. We predict possible dramatic changes
in the force depending on either the host medium, the polarization
and the nature of the surface wave.
\end{abstract}
\maketitle

 ] 

\noindent The expression for the optical force on a small
magnetodielectric particle  \cite{Chaumet_magnet} consist of three
terms: an electric, $<{\bf F}_e>$, a magnetic, $<{\bf F}_m>$,  and
an electric-magnetic dipolar interaction component, $<{\bf
F}_{e-m}>$, whose physical meaning, associated to the differential scattering cross section, was given \cite{Nieto1} on the basis of  the
formal analogy between the conservation of the momentum,
(the optical force), and the energy, (the optical theorem). Studies
on photonic forces in the near field are of importance if one wishes
to enter in the subwavelength scale \cite{ Novotny1, Nieto2,
Nieto3, Nieto6, Quidant1, Quidant2, Kawata, Novotny2,  Song,
Quidant3}. This involves  evanescent waves. However,  as far as we are aware, the effects of
such waves on magnetodielectric objects, have not been yet
investigated.

Like for non-magnetic particles  \cite{MNV&Ricardo, Albaladejo},
each of those above mentioned three terms of the optical force admit
a decomposition into a gradient, a scattering (radiation pressure)
and a curl component. Let ${\bf e}^{(i)} ({\bf s}_{xy})$ and   ${\bf
b}^{(i)}({\bf s}_{xy})$ denote the complex amplitude of each angular
plane wave component of the electric and magnetic vectors incident
on the particle \cite{Mandel&Wolf, MNV1}, inducing  the electric and
magnetic dipole moments ${\bf p}$ and ${\bf m}$, respectively; the
wavevector of this wave being: ${\bf k}=k({\bf s}_{xy}, s_z)$,
$k=n\omega/c$, $\omega$ representing the frequency and
$n=\sqrt{\epsilon \mu}$, ($\epsilon$ and  $\mu$ stand for the
dielectric and magnetic constants of the lossless surrounding
medium),  $s_z=\sqrt{1-s_{xy}^2}$ for propagating waves: $s_{xy}^2  \leq
1$, and $ s_z=\sqrt{s_{xy}^2-1}$ for evanescent waves: $s_{xy}^2 > 1$.
A time dependence $\exp(-i\omega t)$ is assumed throughout.

Let the particle be a small sphere of radius $a$, with constants
$\epsilon_p$ and $\mu_p$, $n_p=\sqrt{\epsilon_p \mu_p }$, such that
its scattering is accurately described (see details in
\cite{Juanjo}) by the electric and magnetic Mie coefficients $a_1$
and $b_1$ \cite{Bohren}. Then the  electric and magnetic
polarizabilities are: $\alpha_e = i \frac{3 \epsilon } {2 k ^3} a_1$
and $\alpha_m = i \frac{3} {2\mu k ^3} b_1$, respectively
\cite{Nieto1}. The induced dipole moments are expressed in terms of
the amplitude of the incident field components: ${\bf p}= \alpha_e
{\bf e}^{(i)}$; ${\bf m}= \alpha_m {\bf b}^{(i)}$. We write
\cite{Nieto1, Draine}: $ \alpha_e = \alpha_e^{(0)} \left(1- i
\frac{2}{3 \epsilon } k ^3 \alpha_e^{(0)} \right)^{-1} ,
\,\,\,\,\,\,\,\, \alpha_m = \alpha_m^{(0)}\left(1- i \frac{2}{3} \mu
k ^3 \alpha_m^{(0)}\right)^{-1} , $ $\alpha_e^{(0)}$ and
$\alpha_m^{(0)}$ being the corresponding static polarizabilities.
Notice that in particular,  in the Rayleigh limit $ka\ll 1$, $k|n_p|
a\ll 1$ one has \cite{Nieto1, Bohren}: $\alpha_e^{(0)}=\epsilon a^3
\frac{\epsilon_p-\epsilon }{\epsilon_p+ 2 \epsilon },
\,\,\,\,\,\,\,\,  \alpha_m^{(0)}=\mu ^{-1} a^3 \frac{\mu_p-\mu
}{\mu_p+ 2 \mu }$. Such magnetodielectric spheres have recently been
shown to be available of $Si$, $Ge$ and $TiO_2$
 in the near infrared \cite{Juanjo}, as well as high refractive index spheres and rods in the microwave
region (see Refs. 17 - 21 in \cite{Juanjo}).

We next determine the action exerted by one of those aforementioned
components: an evanescent wave, on the sphere. Such a wave may be
produced by  e.g. {\it total internal reflection} (TIR) at a plane
dielectric interface, or by a {\it surface plasmon-polariton} (SPP)
at a metal film surface. In either situation, the particle is in the
rarer medium of constants $\epsilon $, $\mu $ and $n =
\sqrt{\epsilon  \mu }$,  in the half space $Z>0$. $X Z$ is the plane
of incidence, (see the inset of Fig.1).  We shall consider the broad
variety of cases, delimited in e.g. \cite{Quidant3,MNV&Ricardo}, in
which both the particle scattering cross section and distance to the
surface allow  to neglect multiple  wave interactions  bettween the
sphere and the plane. The electric and magnetic vectors of the
generic evanescent wave, created in the medium  $Z>0$ containing the
particle, are: ${\bf E}^{(i)}(K)= \left(-\frac{iq}{k } T_{
\parallel}, T_{\perp}, \frac{K}{k } T_{
\parallel}\right) \exp(iKx-qz)$,

${\bf B}^{(i)}(K)= n \left(-\frac{iq}{k } T_{ \perp}, -
T_{\parallel}, \frac{K}{k } T_{ \perp}\right)\exp(iKx-qz)$.
\newline
For TE or $s$ (TM or $p$) - polarization , i.e. ${\bf E}$  (${\bf
B}$) perpendicular to the plane of incidence $XZ$, only those
components with the transmission coefficient $T_{ \perp}$,
($T_{\parallel}$) are chosen in the incident fields ${\bf E}^{(i)}$
and ${\bf B}^{(i)}$ . Now $ K$ denotes the component of the
wavevector ${\bf k}=k({\bf s}_{xy}, s_z)$, parallel to the
interface; i.e. ${\bf k}=( K, 0, iq)$, $q= \sqrt{K^2 -k ^2}$, $k ^2=
K^2-q^2$.

Introducing the above equations into Eqs. (42) - (44) of
\cite{Nieto1}, the resulting electric dipole force components then
are
\begin{eqnarray}
<{\bf F}_e>_{x}= \frac{\Im \alpha_e}{2}K \exp(-2qz)  [|T_{
\perp}|^2 \nonumber \\
+|T_{\parallel}|^2  (2\frac{K^2}{k ^2}-1)], \label{53}\\
 <{\bf F}_e>_{z}= - \frac{\Re \alpha_e}{2}q \exp(-2qz) [
|T_{\perp}|^2   \nonumber \\
+|T_{\parallel}|^2 (2\frac{K^2}{k ^2}-1)]. \label{54}
\end{eqnarray}
Whereas the magnetic dipole forces become
\begin{eqnarray}
<{\bf F}_m>_{x}= n^2 \frac{\Im \alpha_m}{2}K \exp(-2qz) [|T_{
\parallel}|^2 \nonumber\\
+ |T_{\perp}|^2
(2\frac{K^2}{k ^2}-1)],\label{55} \\
 <{\bf F}_m>_{z}= - n^2 \frac{\Re
\alpha_m}{2}q \exp(-2qz) [|T_{
\parallel}|^2 \nonumber\\
+ |T_{\perp}|^2 (2\frac{K^2}{k ^2}-1)].\label{56}
\end{eqnarray}
And the electric-magnetic dipolar interaction forces are given by
\begin{eqnarray}
<{\bf F}_{e-m}>_{x}= -\frac{k ^4}{3}\sqrt{\frac{\mu }{\epsilon }}
\frac{1}{k}\Re (\alpha_e \alpha_m^{*}) K \exp(-2qz)  [|T_{\perp}|^2 \nonumber\\
+ |T_{\parallel}|^2 (2\frac{K^2}{k
^2}-1)],\label{57}\\
 <{\bf F}_{e-m}>_{z}= -
\frac{k ^4}{3}\sqrt{\frac{\mu }{\epsilon }}\frac{1}{k} \Im (\alpha_e
\alpha_m^{*})q \exp(-2qz)[|T_{\perp}|^2 \nonumber\\
+ |T_{\parallel}|^2 (2\frac{K^2}{k ^2}-1)].\label{58}
\end{eqnarray}
$\Re$ and $\Im$ mean real and imaginary parts, respectively. Once
again, for $s$ ($p$) polarization, only the terms with $T_{\perp}$,
($T_{\parallel}$) are taken in these equations. Out of a Mie
resonance, for pure  electric dipole or pure magnetic dipole
particles, with positive $\epsilon_p$ and $\mu_p$  and with little
absorption, $\Re \alpha > \Im \alpha$, and Eqs.
({\ref{53})-({\ref{56}) show that the $Z$-{\it component}, i.e. the
{\it gradient force},  of  $<{\bf F}_{e}>$ and $<{\bf
F}_{m}>$ is  larger than the $X$-{\it component}, i.e. than the {\it
radiation pressure} or {\it scattering force}. This would confirm
some of the observations in the experiment of \cite{Quidant3} for
purely dielectric particles, and predicts an analogous effect in
$<{\bf F}_{m}>$ for magnetodielectric particles. However, this
behavior competes with the opposite one of the interaction force
$<{\bf F}_{e-m}>$,  for which Eqs. ({\ref{57}) and ({\ref{58}) show
that the $X$-component is larger than the $Z$-component. On the
other hand, in a Mie resonance: $\Re \alpha << \Im \alpha$
\cite{MNV&Ricardo,Quidant1,Juanjo} and, hence,  the
opposite effect should occur in its neighborhood,

The term with factor $2 q^2/k ^2=(2 K^2/k ^2)-2$ in the
$X$-component of $<{\bf F}_{e}>$ and $<{\bf F}_{m}>$, Eqs.
(\ref{53}) and (\ref{55}), is due to the {\it curl component} of the
decomposition [cf. Eqs. (42)- (44) in \cite{Nieto1}] of $<{\bf
F}_{e}>$ and $<{\bf F}_{m}>$, respectively. On the other hand, the
contribution of this term to the $Z$-component of $<{\bf F}_{e}>$
and $<{\bf F}_{m}>$, Eqs. (\ref{54}) and (\ref{56}), is totally due
to the gradient component.

For example, for a perfectly conducting Rayleigh sphere:
$\alpha_e^{(0)} = \epsilon  a^3$, $\alpha_m^{(0)} =-a^3/2 \mu$, so
that:
\begin{eqnarray}
<{\bf F}_x>= \frac{K}{2} \exp(-2qz)k ^3 a^6 \epsilon \{
|T_{\perp}|^2[\frac{2}{3}+\frac{1}{6}(2\frac{K^2}{k
^2}-1)\nonumber\\
+\frac{1}{3}]+|T_{\parallel}|^2[\frac{2}{3}(2\frac{K^2}{k
^2}-1)+\frac{1}{6}+
\frac{1}{3}(2\frac{K^2}{k^2}-1)]\}, \label{59}\\
<{\bf F}_z>= -\frac{q}{2} \exp(-2qz) a^3 \epsilon \{
|T_{\perp}|^2[1-\frac{1}{2}(2\frac{K^2}{k
^2}-1)]\nonumber\\
+ |T_{\parallel}|^2[(2\frac{K^2}{k ^2}-1)-\frac{1}{2}]\}.\label{60}
\end{eqnarray}
In the special case of   a particle small enough to neglect terms $k^6 a^6$ and higher, $<{\bf F}_{e-m}>$ does not contribute to $<{\bf F}_z>$.
In Eqs. (\ref{59}) - (\ref{60}) the first and second terms within
the square bracket that multiplies the corresponding $|T|^2$
transmission coefficient, are due to $<{\bf F}_e>$ and $<{\bf
F}_m>$, respectively. The force $<{\bf F}_{e-m}>$ is the third term
in these brackets of Eq. (\ref{59}). The force
$X$-component differs from the $Z$-component by a factor $k ^3
a^3 $.

If the medium containing the particle were a hypothetical
quasi-lossless {\it left-handed} fluid (LHM) \cite{Veselago},
$\epsilon < 0$,  $\mu < 0$  which conveys  $n < 0$, then instead of
$iq$ one has $-iq$ in all equations and, within the regularization
conditions for the wavefunction in $LHMs$ \cite{MNV1}, the forces
exponentially increase with the distance to the surface.

The scattering components of  $<{\bf F}_e>$ and $<{\bf F}_m>$, Eqs.
(\ref{53}) and (\ref{55}),  change sign with that of
$n$, (i.e. with that of $\epsilon$ and $\mu$); thus as regards the
sign of the radiation pressure in LHMs, one obtains the same result
as for an incident  plane propagating wave \cite{Veselago, Nieto1}.
The sign of the scattering component of $<{\bf F}_{e-m}>$,
Eq. (\ref{57}), and of the gradient components, Eqs. (\ref{54}), (\ref{56})   and
(\ref{58}), depends on the relative values of the particle constitutive constants with respect to those
 of the embedding medium. Notice that for the polarizabilities  of
these small particles, a change of sign in $n$, (i.e., in
$\epsilon$ and $\mu$), is equivalent  to a change of sign in $n_p$, (namely, in
$\epsilon_p$ and $\mu_p$).


\begin{figure}[htbp]
\centerline{\includegraphics[width=11cm,clip]{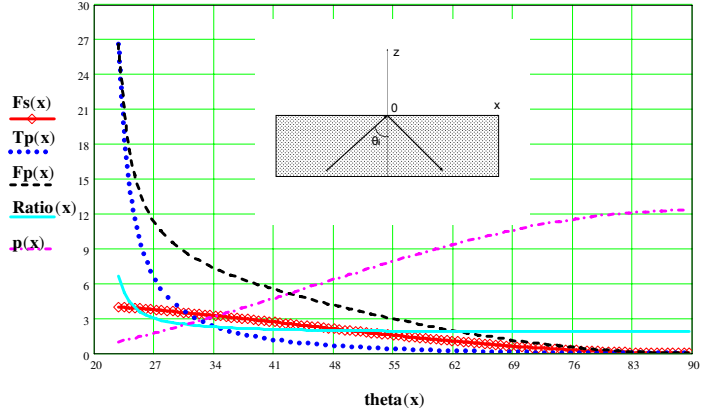}}
\caption[onecolumn]{T.I.R.: With reference to Eqs. (\ref{53})-
(\ref{54}) and (\ref{57}) - (\ref{58}), $F_s(x)$ denotes the force
component factor: $|T_{\perp}|^2$, $F_p(x)$ means the same for the
factor: $|T_{\parallel}|^2 (2\frac{K^2}{k ^2}-1)$, $T_p(x)$ stands
for $|T_{\parallel}|^2$, $Ratio(x)=F_{p}(x)/ F_{s}(x)$,
$p(x)=2\frac{K^2}{k ^2}-1$. $theta(x)$ is the angle of incidence in
degrees, $\theta_i$, from the denser medium of relative index
$n=2.58$ in $Z<0$, (see inset geometry).}
\end{figure}

Returning to the case of ordinary surrounding media, i.e. $n>0$, the factor
$(2(K^2/k ^2)-1)$ associated to the polarization that appears in
Eqs. (\ref{53}) - (\ref{60}), may be large when $K^2>>k ^2$, a
situation that occurs in the {\it electrostatic approximation}. In
TIR, this can be attained either by illuminating at large angles of
incidence, by employing a large contrast interface, or  by a
combination of both. However, these possibilities are hindered by a
consequent drastic decrease of the Fresnel transmission coefficient.
Fig. 1 illustrates these facts for the electric force, Eqs.
(\ref{53}) and (\ref{54}), and for the electric-magnetic interaction
force, Eqs. (\ref{57}) - (\ref{58}), for an index contrast: $2.58$.
Even so, this figure also shows that the ratio $|T_{\parallel}|^2
(2\frac{K^2}{k ^2}-1) / |T_{ \perp}|^2$ can be almost $6$ near the
critical angle $\theta_i=22.8^{\circ}$. These results indicate a
similar interplay of the factor $(2\frac{K^2}{k ^2}-1)$ and  $|T_{
\perp}|^2$   for the magnetic force, Eqs. (\ref{55}) and (\ref{56}).

An  alternative, more efficient than TIR to enhance the optical
force at optical wavelengths with $p$-polarization, and hence with the contribution of the
factor $2(K^2/k ^2)-1$, is when the evanescent wave  is a {\it SPP} emerging at the plane $Z=0$ from a noble
metal film in $Z<0$. For instance, for silver in the
Kretschmann-Raether configuration (cf.
 \cite{Kretch} and Fig. 1(b) of \cite{Raether}), the transmission coefficient of the evanescent wave in $Z>0$
 reaches a value as large as
 $|T_{\parallel}|^2 \simeq 100$ at $\lambda=600 nm$, and   $|T_{\parallel}|^2 \simeq 50$
 at $\lambda=450 nm$ for an angle of incidence on the metal layer (of thickness: $500nm$) from the quartz prism: $\theta_i= {45.2}^{\circ}$,
 [cf. Ref. \cite{Raether}, Section 2.4: Figs. 2.12 and 2.13 and Eqs.  (2.27) - (2.30)], when
the Fabry-Perot resonance of the metal film is
 excited. Then, $2(K^2/k ^2)-1=2\sin^2 \theta_t-1=2 n_0^2 \sin^2 \theta_i-1$,    $\theta_t$ denoting
 the (complex) angle of transmission into the medium $Z>0$ containing the  particle, (which we assume of unity
 refractive index), and  $n_0=2.2$  standing for the quartz refractive
 index. Thus, $2(K^2/k ^2)-1=2 (2.2)^2 \sin^2 45.2 ^{\circ} -1 = 3.87$.
 Namely,  $|T_{\parallel}|^2  (2K^2/k ^2-1)=387$ for $\lambda=600
 nm$ and 193.5 for $\lambda=450 nm$. These values are more than one order of magnitude
 larger than those obtained by TIR, as seen from a comparison with
 Fig. 1.

Concerning the magnetic dipole force on a magnetodielectric
particle, Eqs. (\ref{55}) and (\ref{56}), the  role played by ${\bf E}$ for
the electric force is now played by ${\bf B}$. A strong surface wave
under $s$-polarization, constituting a {\it magnetic SPP}, may now
be excited if the medium in $Z<0$ is magnetodielectric instead of a metal.
 A particular case is that of a left-handed metamaterial in
$Z<0$ \cite{Darmanyan}. Further work, both theoretical and
experimental, is necessary to estimate the strength and feasibility
of such excitation.

In conclusion, we have reported what we believe are the first
predictions on effects of the optical force exerted by an
evanescent, or surface, wave on a magnetodielectric sphere,
immersed in an an arbitrary lossless uniform medium medium.  Since such particles are now known
to be  available \cite{Juanjo}, future experimental work
should be  feasible,  observing these findings and possible additional phenomena.

Work supported by grants of the Spanish MEC: Consolider
\textit{NanoLight} CSD2007-00046, FIS2006-11170-C01-C02
 and FIS2009-13430-C01-C02.


\end{document}